\documentclass[12pt]{iopart}
\usepackage[colorlinks=true,urlcolor=blue,linkcolor=blue]{hyperref}
\usepackage{graphicx}
\usepackage{epsfig}
\usepackage{bm}
\usepackage{iopams}
 \relax

\newcommand{\rld}{\rho_{\Lambda D}}
\newcommand{\rl}{\rho_{\Lambda}}
\newcommand{\rlf}{\rho_{\Lambda 5}}

\begin{document}

\title{Holographic Dark Energy in Braneworld Models with Moving Branes and the $w=-1$ Crossing}

\author{E.~N.~Saridakis
\footnote{E-mail: msaridak@phys.uoa.gr}} \address{Department of
Physics, University of Athens, GR-15771 Athens, Greece}

\begin{abstract}
We apply the bulk holographic dark energy in general 5D two-brane
models. We extract the Friedmann equation on the physical brane
and we show that in the general moving-brane case the effective 4D
holographic dark energy behaves as a quintom for a large
parameter-space area of a simple solution subclass. We find that
$w_\Lambda$ was larger than $-1$ in the past while its present
value is $w_{\Lambda_0}\approx-1.05$, and the phantom bound
$w_\Lambda=-1$ was crossed at $z_{p}\approx0.41$, a result in
agreement with observations. Such a behavior arises naturally,
without the inclusion of special fields or potential terms, but a
fine-tuning between the 4D Planck mass and the brane tension has
to be imposed.
\end{abstract}

 \maketitle

\section{Introduction}

Holographic dark energy \cite{Li,hol1,Guberina,Gong} is a recently
developed ingenious idea of explaining the observed Universe
acceleration \cite{observ}. It arises from the cosmological
application \cite{holcosm} of the more fundamental holographic
principle \cite{Hooft,witten0}. Although there are some objections
about the applicability of holography to a cosmological framework
\cite{Linde}, holographic dark energy has opened new research
directions, revealing the dynamical nature of vacuum energy by
relating it to cosmological volumes. The background on which it is
based, is the black hole thermodynamics \cite{BH,5Dradius} and the
connection between the UV cut-of of a quantum field theory, which
is related to vacuum energy, and a suitable large distance of the
theory \cite{Cohen}. This connection, which was also known from
AdS/CFT correspondence, proves to be necessary for the
applicability of quantum field theory in large distances. The
reason is that while the entropy of a system is proportional to
its volume the black hole entropy is proportional to its area.
Therefore, the total energy of a system should not exceed the mass
of a black hole of the same size, since in this case the system
would collapse to a black hole violating the second law of
thermodynamics. In holographic statistical physics terms this is
equivalent to the exclusion of those degrees of freedom that would
collapse. When this approach is applied to the Universe, the
resulting vacuum energy is identified as holographic dark energy.

Until now, almost all works on the subject have been formulated in
the standard 4D framework. However, brane cosmology, according
which our Universe is a brane embedded in a higher-dimensional
spacetime \cite{Rubakov83,RS99b}, apart from being closer to a
higher-dimensional fundamental theory of nature, it has also great
phenomenological successes \cite{branereview}. In our recent works
\cite{manos.restored} we presented a generalized and restored
holographic dark energy in the braneworld context. The basic
argument was that in a higher-dimensional spacetime, it is the
bulk space which is the natural framework for the cosmological
application (concerning dark energy) of holographic principle, and
not the lower-dimensional brane-Universe. This is obvious since it
is the maximally-dimensional subspace that determines the
properties of quantum-field or gravitational theory (such as
cut-off's and vacuum energy), and this holds even if we consider
brane cosmology as an intermediate limit of an even
higher-dimensional fundamental theory of nature. To be more
specific we recall that in braneworld models, where the spacetime
dimension is more than 4, black holes will in general be
D-dimensional \cite{BH,5Dradius}, no matter what their 4D
effective (mirage) effects could be. Therefore, although
holographic principle is itself applicable to arbitrary dimensions
\cite{Hooft,Verlinde} its cosmological application concerning dark
energy should be considered in the maximal uncompactified space of
the model, i.e. in the bulk. Subsequently, this bulk holographic
dark energy gives rise to an effective 4D dark energy with
``inherited" holographic nature, and this one is present in the
(also arisen from the full dynamics) Friedmann equation of the
brane. One can in general acquire either different or exactly
identical 4D behavior, comparing to that obtained in conventional
4D literature \cite{Li,hol1,Guberina,Gong}. However, even in the
second case, the physical interpretation is radically different.

In our previous works \cite{manos.restored} we used, as a specific
example, a general braneworld model with one brane. The arbitrary
large extra dimension of this case imposes no restrictions on the
application of bulk holographic dark energy, thus we recovered all
the results of 4D literature. In the present paper we are
interested in investigating the case where the bulk is finite, and
for this purpose we use the well explored two-brane model, where
the branes constitute the boundaries of the extra dimension
\cite{RS99b,brcod,twobrane,twobrane2}. From the first moment it
becomes clear that if the branes are steady, i.e. the extra
dimension has a constant size, then the bulk dark energy loses its
dynamical-holographic nature. The reason is that, applying the
holographic dark energy arguments, we cannot consider an arbitrary
large bulk black hole in this case. Although we could still use
non-spherical exotic solutions such as black rings and black
``cigars" \cite{blackring}, or highly rotational or/and charged
black holes \cite{cyllindrical}, if we desire to maintain the
simplicity and universality which lies in the basis of holographic
dark energy we must remain in the aforementioned framework, i.e
preventing the horizon of a spherical bulk black hole being larger
than half the interbrane distance. It seems that holographic dark
energy is in contradiction with brane stabilization mechanism
\cite{Goldberger99}.

In this work we examine braneworld models where the bulk is finite
but with moving boundaries (branes). In this case holographic dark
energy is applicable and the corresponding cosmological length is
the interbrane distance. However, one still needs an additional
fine-tuning assumption, arising form the time variation of the
model parameters. We want to study the behavior of the effective
4D holographic dark energy, and especially its dependence on the
metric scale factor. The rest of the text is organized as follows:
In section \ref{HDEBulk} we present the
 holographic dark energy in the bulk and in section
\ref{HDEBrane} we apply it to a general
 two-brane model in 4+1 dimensions. Finally, in \ref{discussion} we
discuss the physical implications of the obtained results.

\section{Formulation of Holographic Dark Energy in a General Bulk}\label{HDEBulk}

In this section we display the basic results of the bulk
holographic dark energy, formulated in \cite{manos.restored}. The
mass $M_{BH}$ of a spherical and uncharged D-dimensional black
hole is related to its Schwarzschild radius $r_s$ through
\cite{5Dradius,BHTEV}:
\begin{equation}
M_{BH}=r_s^{D-3}
(\sqrt{\pi}M_D)^{D-3}M_D\frac{D-2}{8\Gamma(\frac{D-1}{2})},
\label{5drad}
\end{equation}
where the D-dimensional Planck mass $M_D$ is related to the
D-dimensional gravitational constant $G_D$ and the usual
4-dimensional Planck mass $M_p$ through:
\begin{eqnarray}
M_{D}=G_D^{-\frac{1}{D-2}}, \nonumber\\
M_p^2=M_D^{D-2}V_{D-4},\label{m5m4}
\end{eqnarray}
with $V_{D-4}$ the volume of the extra-dimensional space
\cite{5Dradius}.

If $\rld$ is the bulk vacuum energy, then application of
holographic dark energy in the bulk gives:
\begin{equation}
\rld {\rm{Vol}}({\mathcal{S}}^{D-2})\leq r^{D-3}
(\sqrt{\pi}M_D)^{D-3}M_D\,\frac{D-2}{8\Gamma(\frac{D-1}{2})},
\label{resHDE}
\end{equation}
where ${\rm{Vol}}({\mathcal{S}}^{D-2})$ is the volume of the
maximal hypersphere in a $D$-dimensional spacetime, given from:
\begin{equation}
{\rm{Vol}}({\mathcal{S}}^{D-2})=A_D\,r^{D-1} \label{v2k},
\end{equation}
with
\begin{eqnarray}
A_D=\frac{\pi^{\frac{D-1}{2}}}{\left(\frac{D-1}{2}\right)!}\nonumber,\\
 A_D=\frac{\left(\frac{D-2}{2}\right)!}{\left(D-1\right)!} 2^{D-1}\, \pi^{\frac{D-2}{2}},
\label{ad}
\end{eqnarray}
for $D-1$ being even or odd respectively. Therefore, by saturating
inequality (\ref{resHDE}) introducing $L$ as a suitable large
distance and $c^2$ as a numerical factor, the corresponding vacuum
energy is,  as usual, viewed as holographic dark energy:
\begin{equation}
\rld =c^2 (\sqrt{\pi}M_D)^{D-3}M_D A_D^{-1}
\frac{D-2}{8\Gamma(\frac{D-1}{2})}\,L^{-2} \label{resHD2}.
\end{equation}
As was mentioned in \cite{manos.restored}, the ``suitable large
distance'' which was used in the definition of $L$ in
(\ref{resHD2}) should be the Hubble radius, the particle horizon,
or the future event horizon \cite{Li,Hsu,Gong}, with the last
ansatz being the most appropriate. However, in the case of a
finite bulk with varying size it is obvious that $L$ should be
just this bulk size, with the reason being again the foundations
of holographic dark energy which prevent the use of a length
larger than that. On the other hand, if the future event horizon
is smaller than the bulk size then one should use that instead of
the bulk size. In this case our braneworld model does not ``feel''
the other bulk boundary and the picture is equivalent to the
single-brane model investigated in \cite{manos.restored}.

Finally, let us make a comment concerning the sign of bulk
holographic dark energy. In the original Randall-Sundrum model
\cite{RS99b} the bulk cosmological constant should be negative in
order to acquire the correct localization of low-energy gravity on
the brane. Such a negativity is not a fundamental requirement and
is not necessary in more complex, non-static models, like the one
of the application of the next section. However, as was mentioned
in \cite{manos.restored} and generally speaking, holographic dark
energy is a simple idea of bounding the vacuum energy from above.
It would be a pity if, despite this effort, one could still have a
negative vacuum energy unbounded from below, because then
holographic dark energy would loose its meaning. If holography is
robust then one should reconsider the case of a negative bulk
cosmological constant (although subspaces, such as branes, could
still have negative tensions). Another possibility is to try to
generalize holographic dark energy to negative values, in order to
impose a negative bound. The subject is under investigation.

\section{Holographic Dark Energy in 5D Braneworld models with moving branes}\label{HDEBrane}

We are interested in applying the bulk holographic dark energy in
general 5D braneworld models where the bulk is bounded by two
branes. The inclusion of a second brane is probably the best way
of eliminating possible ``naked'' metric singularities, thus
allowing for more complex and realistic models
\cite{RS99b,brcod,twobrane,twobrane2}. We consider an action of
the form:
\begin{equation}
 S=\int d^4xdy\sqrt{-g}\left(M_5^3R-\rlf\right)+\sum_{i=1,2}\int_{br_i}
 d^4x\sqrt{-\gamma}\,\left({\cal{L}}_{br_i}^{mat}-V_i\right),
\label{action}
\end{equation}
In the first integral $M_5$ is the 5D Planck mass, $\rlf$ is the
bulk cosmological constant which is identified as the bulk
holographic dark energy, and $R$ is the curvature scalar of the 5D
bulk spacetime with metric $g_{AB}$. The second term corresponds
to two ($3+1$)-dimensional branes, which constitute the boundary
of the $5D$ space. $\gamma$ is the determinant of the induced 4D
metric $\gamma_{\alpha\beta}$ on them, $V_i$ stand for the brane
tensions and ${\cal{L}}_{br_i}^{mat}$ is an arbitrary brane matter
content \cite{tetradisbulk1}.

As usual the two branes are taken parallel, $y$ denotes the
coordinate transverse to them and we assume isometry along
3-dimensional $\mathbf{x}$ slices including the branes. For the
metric we choose the conformal gauge \cite{brcod,twobrane}:
\begin{equation}
ds^2=e^{2B(t,y)}\left(-dt^2+dy^2\right)+e^{2A(t,y)}d\mathbf{x}^2.
\label{metric}
\end{equation}
This metric choice, along with the residual gauge freedom
$(t,y)\rightarrow(t',y')$ which preserves the $2D$ conformal form,
allows us to ``fix" the positions of the branes. Without loss of
generality we can locate them at $y=0,1$, having in mind that
their physical distance is encoded in the metric component
$B(t,y)$, and at a specific time it is given by
\cite{brcod,twobrane}:
\begin{equation}
L_5(t)\equiv\int_0^1dy\sqrt{g_{55}}=\int_0^1dy\,e^{B(t,y)},
\label{distance}
\end{equation}
quantity that is invariant under the residual gauge freedom in our
coordinates. The reason we prefer the metric (\ref{metric}),
instead of  the usual form in the literature, is that in the later
case the brane positions are in general time-dependent  and the
various boundary conditions are significantly more complicated.
Thus, our coordinates are preferable for numerical calculations,
despite the loss of simplicity in the definitions of some
quantities. Eventually, the physical interpretation of the results
is independent of the coordinate choice.

The non-trivial 5D Einstein equations consist of two dynamical:
\begin{equation}
\ddot{A}-A''+3\dot{A}^2-3A'^2=\frac{2}{3M_5^3}\,e^{2B}\rlf
\label{eomfulla}
\end{equation}
\begin{equation}
\ddot{B}-B''-3\dot{A}^2+3A'^2=-\frac{1}{3M_5^3}\,e^{2B}\rlf,
\label{eomfullb}
\end{equation}
and two constraint equations:
\begin{equation}
-A'\dot{A}+B'\dot{A}+A'\dot{B}-\dot{A}'=0
\label{constraintsa}
\end{equation}
\begin{equation}
2A'^2-A'B'+A''-\dot{A}^2-\dot{A}\dot{B}=-\frac{1}{3M_5^3}\,e^{2B}\rlf,
\label{constraintsb}
\end{equation}
where primes and dots denote derivatives with respect to $y$ and
$t$ respectively. It is easy to show that the constraints are
preserved by the dynamical equations.

We consider a brane-Universe, which as usual is identified with
the brane at $y=0$, containing a perfect fluid with equation of
state $p=w\rho$ (in the following we omit the index $0$ for the
physical brane quantities and we keep the index $1$ for the ones
on the brane at $y=1$). For the hidden brane we consider for
simplicity just a brane tension, although we could also consider
some matter-field content \cite{manos.param}. The reason we use a
second brane is to eliminate possible ``naked'' metric
singularities. Therefore, assuming $S^1/{\mathbb Z}_2$ symmetry
across each brane we restrict our interest only in the interbrane
space.

Integrating on a small $y$ interval around the branes and using
the boundary terms in the action we obtain the following junctions
(Israel) conditions:
\begin{eqnarray}\
\,[A']_0=-\frac{1}{3M_5^3}\,e^{B_0}(\rho+V)\nonumber\\
\,[B']_0=\frac{1}{3M_5^3}\,e^{B_0}(2\rho+3p-V)\label{junctions0}
\end{eqnarray}
for the physical brane, and
\begin{eqnarray}\
\,[A']_1=\frac{1}{3M_5^3}\,e^{B_1}V_1\nonumber\\
\,[B']_1=\frac{1}{3M_5^3}\,e^{B_1}V_1\label{junctions1}
\end{eqnarray}
for the hidden one, where $B_0\equiv B_0(t)$ and $B_1\equiv
B_1(t)$ are the values of $B(y,t)$ at the branes at $y=0,1$
respectively. In the expressions above we use the following
relations, resulting from $S^1/{\mathbb Z}_2$ symmetry, for the
jump of any function across the branes:
\begin{equation}
[Q']_0=2Q'(0^{+})\ \ ,\ \ \ \ \    [Q']_1=-2Q'(1^{-}). \label{z2}
\end{equation}

Finally, the induced $4D$ metrics of the two (``fixed''-position)
branes in the conformal gauge are simply given by
\begin{equation}
ds^2=-d\tau^2+a^2(\tau)\,d\mathbf{x}^2, \label{4Dmetric}
\end{equation}
with
\begin{equation}
d\tau_i=e^{B_i}dt \label{tau}
\end{equation}
\begin{equation}
a_i=e^{A_i} \label{aa}
\end{equation}
the proper times and scale factors of the two branes ($i=0,1$).
Thus, for the Hubble parameter on the branes we acquire
\begin{equation}
H_i\equiv\frac{1}{a}\frac{da}{d\tau}\Big\vert_i
=e^{-B_i}\dot{A}_i, \label{Hubble}
\end{equation}
which is invariant under residual gauge transformations
\cite{brcod}. As we have mentioned, in the following we omit the
index 0 for the physical brane quantities.

In order to acquire the cosmological evolution on the physical
brane we proceed as follows: Equations
(\ref{eomfulla})-(\ref{constraintsb}) hold for the whole
spacetime, including the branes. In the later case we have to use
the junction conditions (\ref{junctions0}),(\ref{junctions1}) for
the calculation of the first spacial derivatives. Therefore,
eliminating $A''$ form (\ref{eomfulla}) and (\ref{constraintsb}),
and making use of (\ref{junctions0}) and of the Hubble parameter
relation (\ref{Hubble}), we finally obtain:
\begin{equation}
\frac{dH}{d\tau}+2H^2=\frac{1}{3M_5^3}\,\rlf+\frac{1}{36M_5^6}\left[-\rho^2+V\rho-3p(\rho+V)+2V^2\right],
\label{braneevol1}
\end{equation}
where $ H(\tau)$ is the Hubble parameter of the physical brane,
with $\tau$ its proper time. The $\rho^2$ term on the right hand
side of (\ref{braneevol1}) is the usual term present in braneworld
cosmology. Taking the the low-energy limit ($\rho\ll V$) and
knowing that conventional Friedmann equations give:
\begin{equation}
\frac{dH}{d\tau}+2H^2=\frac{4\pi}{3M_p^2}(\rho-3p)+\frac{16\pi}{3M_p^2}\,\rl,
\label{fried}
\end{equation}
with $\rl\equiv\rho_{\Lambda 4}$ the 4D dark energy, it is obvious
that brane evolution coincides with that derived from conventional
4D cosmology if we identify:
\begin{equation}
V=48\pi\frac{M^6_5}{M^2_p}, \label{Vm}
\end{equation}
and
\begin{equation}
\rl=\frac{1}{16\pi}\frac{M_p^2}{M^3_5}\,\rlf+24\pi\frac{M^6_5}{M^2_p}.
\label{rl4}
\end{equation}

Relation (\ref{rl4}) provides the (effective in this
higher-dimensional model) 4D dark energy in terms of the bulk
holographic dark energy, which according to (\ref{resHD2}) is
given by:
\begin{equation}
\rlf=c^2\frac{3}{4\pi}M_5^3L^{-2}\label{rlf}.
\end{equation}
The holographic nature of $\rlf$ is the cause of the holographic
nature of $\rl$. As we have already mentioned, in the two-brane
model examined in the present work, the cosmological length $L$
should be the interbrane distance $L_5$, which is given by
(\ref{distance}). Furthermore, using (\ref{m5m4}) we eliminate the
5D Planck mass $M_5$, in terms of the standard 4D $M_p$, through
$M_5^3=M_p^2/L_5$, since $L_5$ is the volume (size) of the extra
dimension. Note that the varying behavior of the extra-dimension
size will give rise to a varying $M_p$, i.e. varying 4D Newton
constant $G_4$, and this is in general an inevitable consequence
of moving-brane models \cite{varnewton,Guberina}. A detailed
discussion on this point is given at the end of this section.
Thus, we finally acquire the following form for the effective 4D
holographic dark energy:
\begin{equation}
\rl=\left(\frac{3c^2}{64\pi^2}+24\pi\right)M_p^2\,L^{-2}_5.
\label{rl4b}
\end{equation}
As we see, we have resulted in a simple holographic relation,
despite the complicated nature of the model.

Our goal is to reveal the dependence of $\rl$ on the physical
brane scale factor $a$. Since we have recovered the standard
Friedmann equation (through identifications (\ref{Vm}) and
(\ref{rl4})) and using (\ref{distance}) for the calculation of
$L_5$ we finally obtain:
\begin{equation}
H^2=\frac{8\pi}{3M_p^2}\,\rho+\frac{8\pi}{3}\left(\frac{3c^2}{64\pi^2}+24\pi\right)
\left(\int_0^1dy\,e^{B(t,y)}\right)^{-2}. \label{fried2}
\end{equation}
Therefore, despite the simple form of relation (\ref{rl4b}),
complexity has reappeared in the non-localized nature of the above
equation. Such a behavior was expected and is a result of the
``global'' properties of bulk holographic dark energy. This is a
radical difference comparing to the single-brane case of
\cite{manos.restored} where the arbitrary large bulk allowed for a
brane-localized equation form. In the present case, the solution
of the full 5D equations is indispensable. Namely, we have to
solve (\ref{eomfulla})-(\ref{constraintsb}) under boundary
conditions (\ref{junctions0})-(\ref{junctions1}), imposing the
conventional time evolution for $\rho$ and $p$. Knowing $A(t,y)$
and $B(t,y)$ we calculate the interbrane distance $L_5(t)$ through
(\ref{distance}) and then $\rl(t)$ through (\ref{rl4b}). We
calculate the physical brane scale factor $a(t)$ using (\ref{aa}),
and by eliminating $t$ we obtain the questioning relation for
$\rl(a)$. Finally, as usual, we identify $\rl(a)$ with $\rl(a)\sim
a^{-3(1+w_\Lambda)}$, and we extract the form of $w_\Lambda(z)$,
with $z=\frac{a_0}{a}-1$ and $a_0$ the value of $a$ at present
time.

The aforementioned procedure for the derivation of $\rl(a)$
relation is impossible to be completed analytically (analytical
solutions can be obtained only for stationary cases, i.e. with
constant interbrane distance, which as we have mentioned are in
contradiction with the holographic nature of dark energy).
However, avoiding a full numerical approach, for the purpose of
this work we examine the following solution class:
\begin{equation}
A(t_0,y)=B(t_0,y)=\ln\left[-\frac{y+q_1}{q_2}\right] \label{sol}
\end{equation}
\begin{equation}
\dot{A}(t_0,y)=\dot{B}(t_0,y)=q_3\left[y+q_1\right],\label{sold}
\end{equation}
which satisfy the constraint equations
(\ref{constraintsa}),(\ref{constraintsb}) at $t_0$ (and therefore
at every $t$ since (\ref{constraintsa}),(\ref{constraintsb}) are
preserved by the equations of motion) provided that
$(1+2q_1)^2q_3^2=c^2/(2\pi)$. Moreover, boundary conditions
(\ref{junctions0}) and (\ref{junctions1}) are fulfilled imposing
$V=6M_5^3q_2/q_1^2$ and $V_1=-6M_5^3q_2/(q_1+1)^2$.

\begin{figure}[h]
\begin{center}
\mbox{\epsfig{figure=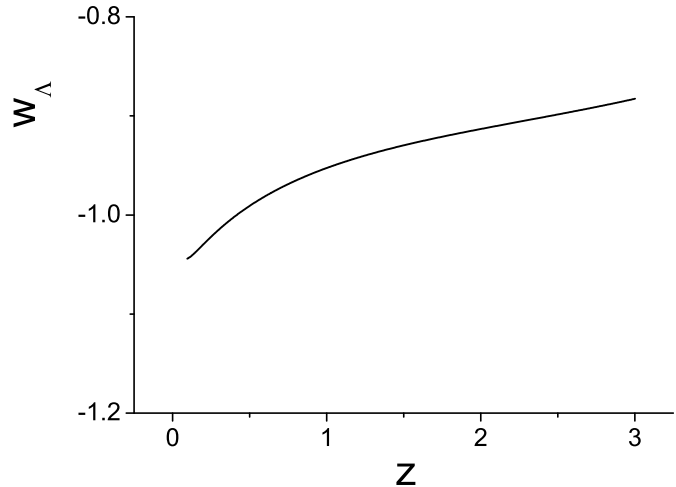,width=9cm,angle=0}} \caption{\it
$w_\Lambda$ versus $z$ for solution class
(\ref{sol})-(\ref{sold}), using $q_1=-1.4$, $q_2\approx11$,
$q_3\approx0.22$, $c=1$ and the unit $M_5=1$. The present value of
$w_\Lambda$ is $w_{\Lambda_0}\approx-1.05$ and the phantom bound
$w_\Lambda=-1$ was crossed at $z_{p}\approx0.41$.} \label{wz}
\end{center}
\end{figure}

Investigating the low-energy (late-time) evolution of the
aforementioned solution class, i.e. omitting $\rho$ and $p$ in
boundary conditions which make them significantly simpler, we
obtain interesting results. In particular, for a large area of the
parameter-space we find a reasonable $w_\Lambda(z)$ form, with the
basic requirement (necessary but not always efficient) being the
decreasing of interbrane distance. Fortunately, a decreasing
interbrane distance seems to have larger probability than an
increasing one (since in the later case the system is often
unstable or ``naked'' singularities do appear between the branes).
Furthermore, there are many parameters present in the model and in
solution class (\ref{sol})-(\ref{sold}). These characteristics
make it substantially easier to acquire a reasonable
$w_\Lambda(z)$ form in the 5D framework described in this work,
than in conventional 4D holographic dark energy
\cite{Li,hol1,Guberina,Gong}. In addition, we do not have to
fine-tune the constant $c$ in the definition of $\rlf$ in relation
(\ref{rlf}). In fig.~\ref{wz} we depict $w_\Lambda(z)$ for
$q_1=-1.4$, $q_2\approx11$, $q_3\approx0.22$ and $c=1$, and using
the unit $M_5=1$ (the specific scale does not affect the
$w_\Lambda(z)$ form).
We observe that the effective 4D holographic dark energy behaves
like a ``quintom'' \cite{quintom}, that is $w_\Lambda$ was larger
than $-1$ in the past while its present value is
$w_{\Lambda_0}\approx-1.05$, and the phantom bound $w_\Lambda=-1$
was crossed at $z_{p}\approx0.41$. This result is in agreement
with late acceleration of the Universe and dark energy constraints
imposed by observations \cite{observHDE1,observHDE}. Note that
quintom behavior arises naturally in our higher-dimensional brane
model, without the inclusion of extra fields or specific potential
terms by hand. Moreover, as we mentioned above, we do not have to
fine-tune the solution parameters, since a large area of the
parameter-space leads to similar behavior (the case of
fig.~\ref{wz} just corresponds  to a good representative in
comparison with observations). Finally, other parameter areas in
the aforementioned solution subclass, as well as a general
numerical investigation beyond this ansatz (a hard task due to
various instabilities \cite{brcod,twobrane}) reveal interesting
but likely un-physical $w_\Lambda$ behavior, such as chaotic
oscillations with respect to $z$. However, this could still be the
case in our Universe for larger $z$. The subject is under
investigation.

Let us finish this section with some comments on the dynamical
nature of some quantities. As we have mentioned in the
introduction, the main goal of the present work is the study of
bulk holographic dark energy in a finite braneworld model. From
the first moment we educe that if the bulk boundaries (the two
branes) are steady then holographic dark energy becomes a
constant, losing its dynamical nature. In other words, the concept
of holographic dark energy is in contradiction with finite-bulk
framework, unless we consider moving-brane models as a necessary
outlet. Unfortunately, this choice leads to some undesirable
consequences. Indeed, with a varying interbrane distance it is
obvious from (\ref{m5m4}) that either the 5D Planck mass $M_5$ or
its 4D counterpart $M_p$, or even both, should change with time.
Since $M_5$ is a fundamental quantity of our model we desire to
maintain it as a constant. Thus, $M_p$ and therefore the 4D
Newton's gravitational constant $G_4$ are the ones that reflect
the dynamical nature of the bulk, and this is a common feature of
moving-brane models \cite{varnewton,Guberina}. However, in order
to acquire the correct effective 4D cosmological evolution we need
to make the identification (\ref{Vm}), a usual approach of
braneworld cosmology. This relation ``matches'' the 4D Planck mass
(which is determined by dimensional reduction of the 5D action)
with the brane tension (which should be given by the vacuum-energy
predicted by the effective 4D QFT), and brane-cosmologists hope to
acquire its justification by a fundamental theory of nature,
unknown up to now. Unfortunately, in the present model of moving
branes this fine-tuning must hold at all times and this is
definitely an additional assumption, independent of the rest
formulation. Thus, the extension of holographic dark energy to
finite-bulk models is still obscure, since one expects from a
fundamental theory to justify the aforementioned eternal
fine-tuning.

\begin{figure}[h]
\begin{center}
\mbox{\epsfig{figure=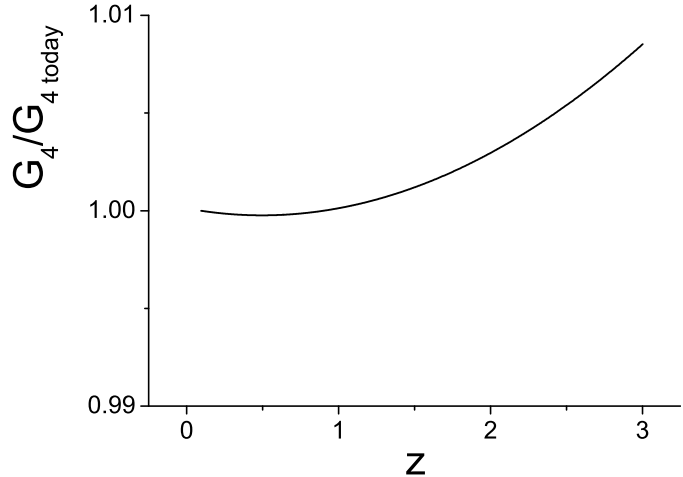,width=9cm,angle=0}} \caption{\it
Evolution of the 4D Newton's constant $G_4$, divided by its
present value, versus $z$ for solution class
(\ref{sol})-(\ref{sold}), using $q_1=-1.4$, $q_2\approx11$,
$q_3\approx0.22$, $c=1$ and the unit $M_5=1$. The minimum value is
obtained at $z_{p}\approx0.41$, which corresponds to the
phantom-divide crossing of $w_\Lambda(z)$.} \label{G44}
\end{center}
\end{figure}

In order to provide a clear picture of these features, in
figure~\ref{G44} we depict the evolution of the 4D Newton's
constant $G_4$ divided by its present value, in terms of $z$, for
the same parameter values of fig.~\ref{wz}. We observe that $G_4$
acquires its minimum value at $z_{p}\approx0.41$, which
corresponds to the phantom-divide crossing of $w_\Lambda(z)$.
This behavior of $G_4$ is consistent with cosmological constraints
which restrict its deviation between $\pm5\%$ \cite{G4com}.
However, we mention here that not all numerical solutions satisfy
these limits. Indeed, our numerical elaboration reveals that out
of $10^3$ solutions that present an acceptable (quintom) form for
$w_\Lambda(z)$ (similar to that depicted in fig.~\ref{wz}), only
$\approx 20\%$ correspond to an acceptable $G_4$-behavior, too.

\section{Discussion-Conclusions}\label{discussion}

In this work we apply the bulk holographic dark energy in a
general braneworld model with moving branes. Such a generalized
bulk version of holographic dark energy is necessary if we desire
to match the successes of brane cosmology in both theoretical and
phenomenological-observational level, with the successful, simple,
and inspired by first principles, notion of holographic dark
energy in conventional 4D cosmology. In particular, as we showed
in \cite{manos.restored}, the bulk space is the natural framework
for the cosmological application, concerning dark energy, of
holographic principle, since it is the maximally-dimensional
subspace that determines the properties of quantum-field and
gravitational theory, and the black hole formation. Subsequently,
this bulk holographic dark energy will give rise to an effective
4D dark energy with ``inherited" holographic nature, and this one
will be present in the effective Friedmann equation of the brane.

Applying the bulk holographic dark energy in the well investigated
two-brane model, and using the interbrane distance (bulk size) as
the cosmological length in its definition, we deduce that the
branes have to move in order for dark energy to preserve its
holographic-dynamical nature. In this case we extract the
effective Friedmann equation on the physical brane, which has a
non-localized (on the brane) form since the interbrane distance
must be calculated from the full 5D dynamics. This complexity is a
result of the ``global'' characteristics of bulk holographic dark
energy. Numerical investigation on a simple solution subclass
reveals a quintom behavior \cite{quintom} for $w_\Lambda(z)$. In
particular, $w_\Lambda$ was larger than $-1$ in the past, it
crossed the phantom divide $w_\Lambda=-1$ at $z_{p}\approx0.41$,
and its present value is $w_{\Lambda_0}\approx-1.05$. This
behavior is in notable agreement with observations
\cite{observHDE1,observHDE} which give
$w_{\Lambda_0}=-1.02\pm^{0.13}_{0.19}$ and $z_{p}=0.46\pm0.13$.
However, although we have not included any special fields or
specific potentials, an additional assumption has to be imposed in
order for this holographic dark energy application to be valid.
Namely, the fine-tuning between the brane-tension and the 4D
Planck mass. Furthermore, Newton's constant acquires a dynamical
nature too and one has to be careful in order to be consistent
with cosmological constraints \cite{G4com}. In conclusion, the
investigation of the extension of holographic dark energy in
finite-bulk models reveals that a reasonable
$w_\Lambda(z)$-behavior is acquired relatively easily, only under
a fundamental fine-tuning. Definitely, the combination of
holographic dark energy with brane cosmology is an interesting
subject which needs further investigation.
\\

\section*{{\bf{Acknowledgements:}}}
The author is grateful to  G.~Kofinas, K.~Tamvakis, N. Tetradis,
F. Belgiorno, B. Brown, S. Cacciatori, M. Cadoni, R.~Casadio,
G.~Felder, A.~Frolov, B. Harms, N.~Mohammedi, M.~Setare,
Y.~Shtanov and to an anonymous referee, for useful discussions and
suggestions.

\end{document}